# Mechanism map and genetic analysis of avalanche vulnerable area based on multi-source data

Student: Zhou Zexuan Ma Bing Qi Zhu Jianwei Supervisor: Kang Zhizhong

(College of Land Science and Technology)

**Abstract**: Avalanche disaster is a major natural disaster that seriously threatens the national infrastructure and personnel's life safety. For a long time, the research of avalanche disaster prediction in the world is insufficient, there are only some basic models and basic conditions of occurrence, and there is no long series and wide range of avalanche disaster prediction products. Based on 7 different bands and different types of multi-source remote sensing data, this study combined with existing avalanche occurrence models, field investigation and statistical data to analyze the causes of avalanche. The U-net convolutional neural network and threshold analysis were used to extract the distribution of long time series avalanch-prone areas in two study areas, Heiluogou in Sichuan Province and along the Zangpo River in Palong, Tibet Autonomous Region. In addition, the relationship between earthquake magnitude and spatial distribution and avalanche occurrence is also analyzed in this study. This study will also continue to build a prior knowledge base of avalanche occurrence conditions, improve the prediction accuracy of the two methods, and produce products in long time series interannual avalanch-prone areas in southwest China, including Sichuan Province, Yunnan Province, and Tibet Autonomous Region. The resulting products will provide high-precision avalanche prediction and safety assurance for engineering construction and mountaineering activities in Southwest China.

【 Key words 】 Avalanche U-net Convolutional neural network threshold analysis multi-source remote sensing data

## 1. Research background

All glaciers in China are mountain glaciers, most of which are above 5,000 m above sea level, 80% of which are continental glaciers. Continental glaciers are mainly distributed in the Altai Mountains, the western section of the Tianshan Mountains, the Karakoram Mountains, the western and middle slopes of the Himalayas, the western section of the Nianqing Tanggula Mountains Jiali, the eastern section of the Tanggula Mountains, the Bayankhara Mountains, the Anyimachen Mountains and the eastern section of the Qilian Mountains. Marine glaciers are mainly distributed in the eastern part of Nianqing Tanggula, the Hengduan Mountains in western Sichuan and northern Yunnan, and the eastern part of the Himalayas. According to statistics, there are a large number of avalanches in China's glaciers every year, which is a country with high incidence of avalanche disasters, among which a large number of Marine glaciers are distributed in western Sichuan, northern Yunnan and southeast Tibet, and the avalanche is the most serious, which is the world's avalanche disaster area.

Avalanches are a serious threat to mountaineers. In the danger encountered in the mountain exploration, the harm caused by avalanches is the most frequent and tragic, often resulting in the "entire army" of the mountain, because of the avalanche of people to account for all mountain casualties 1/2 ~ 1/3. Avalanches also have significant damage to infrastructure, in the process of Sichuan-Tibet railway construction, there have been many cases of construction interruption and even missing construction personnel because of avalanches. For the completed construction facilities, there are also many damaged and destroyed because of avalanches, resulting in huge property losses and casualties.

Therefore, the significance of predicting avalanches is particularly significant. In this study, two methods, U-net convolutional neural network and threshold extraction and analysis, were used to combine 7 different bands and different types of multi-source remote sensing data and long time series statistical data, and the prior knowledge of the pregnant conditions of avalanche disasters was established by avalanche occurrence model, field investigation and data statistical methods. The two methods were used to extract the avalanch-prone areas of long time series in winter and summer in the two research areas along the Hailuogou and Palong Zangbo River, and the visual consistency test was carried out on the vulnerable avalanche areas identified by the two methods at the same time in the same area, and the accuracy of the two methods was judged, and the spatial relationship between earthquake and avalanche disaster was further analyzed. Next, this study will use the two methods to produce the avalanche disaster prone areas in southwest China, establish facilities and systems to prevent avalanche disasters, strengthen the emergency response ability, and continue to strengthen the research and prediction of avalanches.

First of all, it can provide guidance for alpine exploration activities. After learning about some avalanch-prone glaciers, the climbing plan can be adjusted according to the actual situation, avoid these glaciers, find other glacier routes less prone to avalanches to climb, or find a time on the route where glaciers are less prone to avalanches to climb. After understanding the conditions under which avalanches occur, you can prepare ahead of time during the climb to try to avoid these risks. Through these two points, the predictability of disasters in alpine exploration activities can be greatly improved, and the risk can be greatly reduced.

Secondly, it can also help with the country's infrastructure. At present, during the period of the 14th Five-Year Plan, the state has invested a lot of manpower, material and financial resources in infrastructure construction. In the infrastructure construction of Sichuan-Tibet glacier area, by understanding the spatial distribution of some avalanch-prone glaciers in advance, we can better plan the infrastructure construction program to avoid these dangerous glaciers. At the same time, mastering the conditions of avalanche can also take these conditions into account in the process of infrastructure construction, so as to avoid the occurrence of accidents. Through these two points, the safety and efficiency of infrastructure projects can be improved as much as possible, so as to make contributions to the national "14th Five-Year Plan" infrastructure construction.

2. Research methods

## 2.1 Overview of the study area

In order to obtain enough avalanche disaster data, the study area must meet the following characteristics: 1. There are a large number of permanent glaciers distributed, and the types of glaciers are mostly Marine glaciers; 2. There are human activities in the area near the glacier all the year round, and there are a lot of data records of avalanche disasters; 3. There are records of long time series of all multi-source remote sensing data needed for research, with typical climate types; 4. It is an area where mountaineering activities and engineering construction are more involved, and has great reference value.

According to the above characteristics, two research areas were selected: the Hailuogou Research area in Garze Tibetan Autonomous Prefecture, Sichuan Province (101.77°E-102.22°E, 29.45°N-30.07°N enclosed area) and the Palong Zangpo River Coastal Research area in Nyingchi City, Tibet Autonomous Region (93°E-98°E, 28.33°N-31°N enclosed area).

Hailuogou research area is a relatively concentrated area of Marine glacier development. The most extensive is Hailuogou Glacier, which covers an area of about 10.6 square kilometers, while the other glaciers are relatively small. These glaciers provide rich glacial landforms and are of great scientific research value. Hailuogou Research area is an alpine zone. Due to its high altitude and complex terrain, the local climate has unique characteristics. The annual average temperature is about -2°C, the lowest temperature in winter can reach about -30°C, and the highest temperature can reach 15°C. At the same time, there is a lot of rain and snow in this area, and the annual precipitation is about 700 mm. The annual snowfall in this area is large, and it is one of the main avalanches in the country. Therefore, in order to prevent avalanche disasters, the region has a relatively complete data recording and monitoring system. The relevant departments will carry out regular monitoring and early warning for the area affected by avalanches, and establish a set of scientific measures and emergency plans to prevent avalanche disasters.

The glaciers along the Palong Zangbo River are widely distributed. They are mainly concentrated in the Himalayas and Nagqu region. The Palong Zangbo River basin is home to a large number of glaciers. There are more than 5,000 existing glaciers in the region, covering an area of more than 7,500 square kilometers. The region along the Palong Zangbo River belongs to the plateau monsoon climate and the alpine subcold climate. The climate is dry, the precipitation is uneven, the temperature difference is large, and the temperature difference between day and night is also very obvious. The average annual temperature in this region ranges from -2°C to 8°C, with the lowest temperature reaching -30°C and the highest temperature reaching 25°C. Precipitation is mainly concentrated in summer, with an average annual precipitation of about 250 mm. The area along the Palong Zangpo River is one of the most active avalanche areas in China, and the avalanche disasters in this area are frequent and large in scale. In recent years, with the advancement of technology, the local avalanche monitoring and early warning system has been greatly developed, and there are more avalanche data records.

At the same time, these two research areas are also the necessary areas for the construction of Sichuan-Tibet Railway, and there are more engineering construction; In addition, there are more typical snow mountains and glaciers such as Nama Peak, Jinjin Mountain and Laigu Glacier along the Gongga Mountain and Palong Zangbo River, which are common places for mountaineering and glacier exploration activities. Therefore, the study of these two areas has typical guiding significance for engineering construction and mountaineering activities.

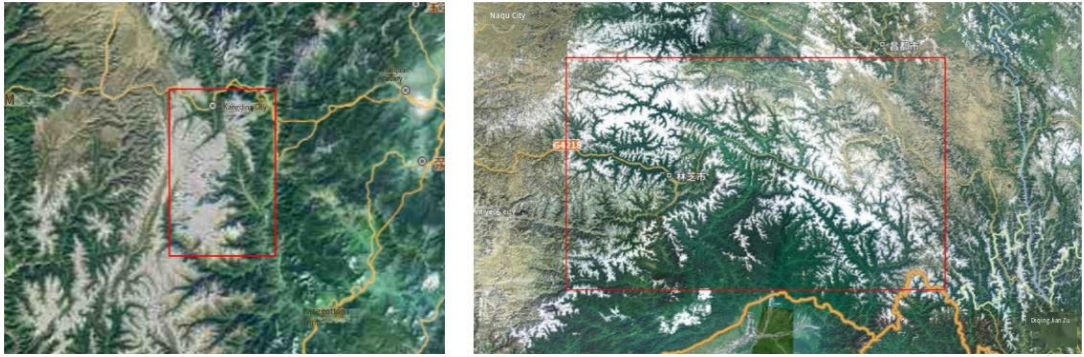

FIG. 1 Schematic diagram of Hailuogou and Palong Zangpo River coastal research area

## 2.2 Multi-source remote sensing data

### 2.2.1 Gaofen-3 SAR L1A data product

Gaofen-3 satellite is a remote sensing satellite of China's Gaofen-3 special project. It is a 1-meter resolution radar remote sensing satellite, and China's first C-band multipolarization synthetic aperture radar (SAR) imaging satellite with a resolution of 1 meter

L1A SLC (Single view complex product) data is a slant range complex data product obtained after image processing and relative radiation correction. In order to improve the visual effect of the image and improve the estimation accuracy of each pixel backscatter, multi-view processing is required, that is, multiple independent samples of the target are aversely superimposed. On the one hand, multi-view processing makes the geometric features of the image closer to the actual situation on the ground, and on the other hand, it also reduces the speckle noise to a certain extent.

### 2.2.2 SRTM 30m Global DEM

The Space Shuttle Radar Topographic Mapping Mission (SRTM) data was jointly measured by NASA and the National Surveying and Mapping Agency (NIMA) of the Department of Defense, and the acquired radar image data was processed over two years to produce a digital terrain elevation model.

The SRTM 30 m DEM dataset is an important raw data for the study and analysis of terrain, watershed and ground object identification. It can reflect local terrain features with a certain resolution. A large amount of surface morphology information can be extracted through it, including the slope and slope direction of the watershed grid units and the relationship between cells. It is widely used in surveying and mapping, hydrology, meteorology, geomorphology, geology, soil, engineering construction, communication, military and other national economy and national defense construction, as well as humanities and natural sciences.

### 2.2.3 China's 30-meter annual land cover products

The annual China Land Cover Dataset (CLCD) was developed by Professor Huang Xin's team of Wuhan University. It was based on 300,000 Landsat images, combined with automatic stabilization samples of existing products and visual interpretation samples. It is divided into 9 land cover types: farmland, forest, shrub, grassland, water body, ice and snow, wasteland, impervious layer and wetland.

The dataset was based on 5,463 independent reference samples, and the overall accuracy of the product was 79.31 percent. The dataset reflects China's rapid urbanization process and a series of ecological projects, revealing the impact of human activities on regional land cover under climate change.

**2.2.4 China's monthly precipitation dataset**

The monthly precipitation dataset of China (1901-2021) was generated in China by Delta spatial downscaling scheme based on the global 0.5° climate dataset released by CRU (Climate Research Institute, University of East Anglia, UK) and the global high-resolution climate dataset released by WorldClim platform. Data from 496 independent meteorological observation points were used for verification, and the verification results were credible. The spatial resolution of the data set is about 1km, and the precipitation unit is 0.1mm. It includes the main land of the country (including Hong Kong, Macao and Taiwan), excluding the islands and reefs in the South China Sea.

**2.2.5 Global surface temperature/emissivity 8-day synthetic data set**

MOD11A2_V6 (MODIS/Terra Land Surface Temperature/Emissivity 8-Day L3 Global 1km SIN Grid) Global surface temperature /Emissivity 8-day composite data set released by NASA, Is a Level 3 product of MODIS and belongs to the land feature product. Its pixel value is the average of 8-day MOD11A1 land surface temperature with a resolution of 1km. The product corrects the edge distortion generated by the imaging process of the remote sensor on the basis of 1B data. The MOD11A2_V6 provides day and night surface temperatures, related quality indicators, observation time, zenith Angle, clear sky coverage, and MODIS bands 31 and 32.

**2.2.6 Statistical data on the frequency and spatial distribution of avalanche disasters in the two study areas during 1980-2020**

The Hengduan Mountain Glacier Basic information dataset of the National Qinghai-Tibet Plateau Scientific Data Center is a collection of statistics on the glaciers and their types in Hengduan Mountain and the information of each glacier, as well as the snow line data and related parameters of some glaciers in China, including the AAR value of some glaciers in Gongga Mountain and the avalanche area (measured data).

The 1:250,000 major engineering disturbance disaster data of the Qinghai-Tibet Plateau of the National Qinghai-Tibet Plateau Scientific Data Center is mainly interpreted by Google earth for disturbance disasters, and combined with field investigation to verify the interpretation results, ArcGIS is used to generate disaster distribution maps from 1985 to 2020, including avalanche and landslide disasters. The original data is of high precision, and the disaster files are generated in strict accordance with the interpretation specifications and reviewed by special personnel, and the data quality is reliable.

According to the above two data combined with avalanche disaster news materials from 1980 to 2020, the number and spatial distribution of avalanche disasters in the study area are calculated.

**2.2.7 Statistical data of earthquake magnitude and spatial distribution in Hailuogou Research Area from 2000 to 2020**

A total of 562 earthquakes with magnitudes above magnitudes 3 around the epicenter were selected from the China Earthquake Network Quick Report catalog database in the historical earthquake data set of magnitudes above magnitudes 2 of the M6.8 earthquake in Lading, Sichuan

Province, from 2009 to the present. The data includes the longitude and latitude of the earthquake, the magnitude of the earthquake and the focal depth.

The Catalogue of Destructive earthquakes on the Tibetan Plateau since 1970, compiled by the National Tibetan Plateau Scientific Data Center, records information such as the epicenter location, time of occurrence and magnitude of 2,854 earthquakes measuring M≥4.7 that have occurred on the Tibetan Plateau and nearby areas since 1970.

According to the above data and the news materials about the earthquakes in Hailuogou area from 2000 to 2020, the magnitude and spatial distribution of the earthquakes in the Hailuogou research area from 2000 to 2020 are calculated.

**2.3 Data processing methods**

The above data are classified and processed in this study, which is mainly divided into two methods to extract avalanch-prone areas. The first method is deep learning extraction based on U-net convolutional neural network. The data used in this method is Gaofen-3 SAR L1A data product, and U-net convolutional neural network is trained to identify avalanch-prone areas. The second method is based on threshold analysis and extraction of avalanche prone conditions. The data used in this method are SRTM 30m global DEM, China's 30-meter annual land cover product, China's monthly precipitation dataset, global surface temperature/emissivity 8-day composite dataset, and statistical data of the number and spatial distribution of avalanche disasters in the two research areas from 1980 to 2020. According to the analysis of the physical model of avalanche formation, the analysis of the threshold conditions of slope, slope direction, precipitation and surface temperature, as well as the field investigation results, the threshold conditions of avalanche susceptibility are obtained, and the avalanch-prone area is extracted according to the threshold conditions.

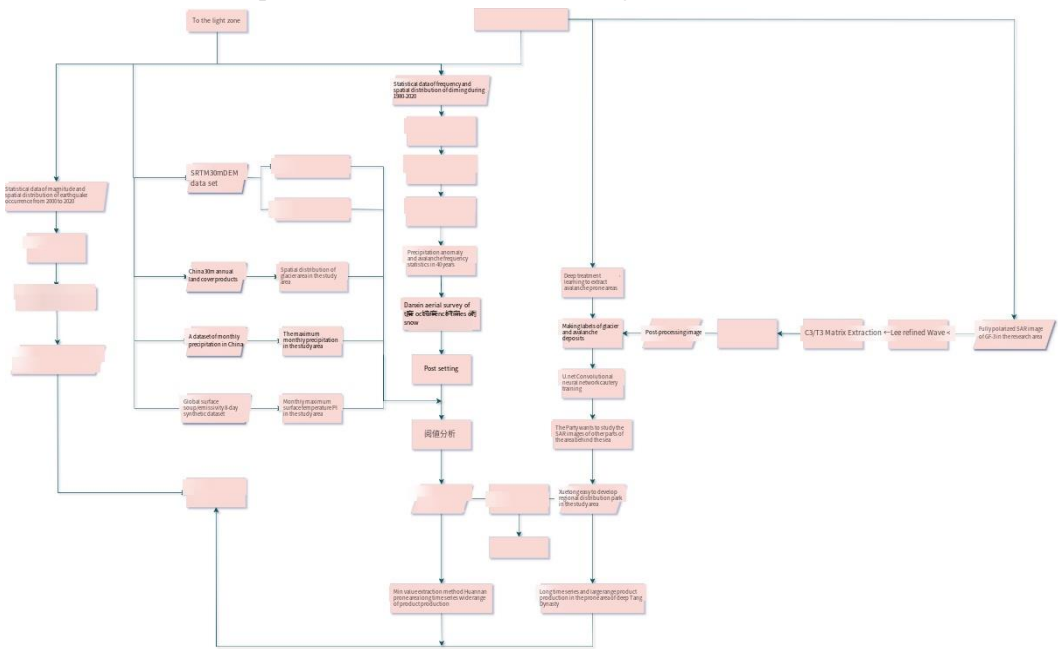

FIG. 2 Research flow chart

**2.3.1 U-net Convolutional neural network extraction**

First, the Gaofen-3 SAR L1A data product in the Palong Zangbo River research area is pre-processed. The PolSARpro software has refined it to three steps: Lee refined filtering, C3/T3 matrix extraction and geometric correction. Then the false-color synthesis is performed to obtain the pre-processed image and perform visual interpretation. Among them, the glacier is a bright white strip with smooth surface and clear texture. The vulnerable avalanche area at the ridge is grayish black, inverted triangle cone, and striped texture. There are more moraines, so the surface is rough, and it is obviously distributed in the river valley and glacier movement channel or on both sides of most glaciers. The labels of glacier and avalanche vulnerable areas were made according to the visual interpretation results, and the labels were divided into training data set, test data set and verification data set according to the proportion of 70%, 10% and 20%. The training data set was put into the constructed U-net convolutional neural network, and a total of 100 iterations were carried out. After the 60th iteration, the neural network fitted well. The final training error is about 0.25 and verification error is about 0.53, indicating that the trained U-net convolutional neural network can accurately identify the distribution area of glacier and avalanche vulnerable area, and can be used to produce products in a long series and a large range of avalanch-prone areas.

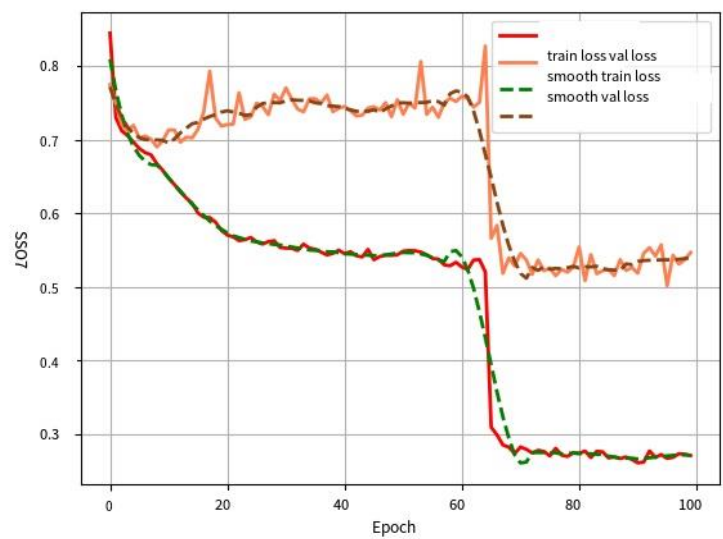

FIG. 3 U-net Convolutional neural network fitting curve

**2.3.2 Analysis of avalanche disaster conditions**

Cui Peng et al. (2018) believe that there are three basic conditions for the development of avalanche disasters in mountain environments, which are as follows: Energy condition, material condition and excitation condition, in which the energy condition is mainly solar radiation, the material condition is mainly a large amount of snow in a short period of time, and the excitation condition is mainly a rise in surface temperature in a short period of time, other related natural disasters, changes in the internal structure of snow layer and human activities. Schweizer et al. (2015) believe that the main influencing factors of avalanches include topography, meteorological conditions and characteristics of snow cover, etc. Topography and geomorphology mainly include slope inclination, slope direction, curvature, distance from ridge, forest cover, etc. Meteorological conditions mainly include precipitation (new snow or rain), wind, temperature, daily radiation, etc. The characteristics of snow cover mainly include weak interlayers of snow cover, etc. The onset of avalanches is usually related to excitation conditions such as strong snowfall, steep rise in temperature, earthquake and human activities. Mcclung et al. (2013) proposed an avalanche disaster model, arguing that heavy snowfall can rapidly increase the snow thickness on the hillside. When the snow thickness is greater than the limit thickness of the snow on the hillside, the cohesion of the snow cannot resist the sliding force formed by gravity traction, and it will slide downward, causing the snow body to collapse. When the snowboard thickness reaches 20 cm, the avalanche of natural release may occur. Podolskiy et al. (2010) proposed the relationship between the upper limit curve of the epicenter distance of earthquake-induced avalanches.

On July 21, 2022 and July 28, 2022, field visits were made to the Hailuogou research area and the Palong Zangbo River coastal research area, respectively. The slope, slope direction and soil temperature formed in the avalanche vulnerable area were measured by compass, GPS receiver, thermometer and other instruments. The expedition was also close to the avalanche vulnerable area. The soil conditions in the lower part of the accumulation body were observed. The local residents were visited on the spot to obtain the information of avalanche disaster.

In Hailuogou research area, it can be observed that there are obvious avalanche vulnerable areas near Gongga Mountain and the surrounding Weifeng, which are mostly distributed in the gullies formed by glacier erosion and glacier meltwater erosion. Most of the vulnerable avalanche areas are triangular cones from top to bottom, with different soil quality above and below, and basically no vegetation cover in their locations. Most of them are distributed southeast, and the average slope is about 45°~70° visually. Observation five times, the average temperature of the soil is about 20.5 degrees Celsius, the soil is relatively dry.

In the study area along the Palong Zangbo River, a large number of avalanche vulnerable areas can be observed distributed on both sides of the Palong Zangbo River canyon, formed by the downward development of glaciers on snow-covered mountains, and many remnants of avalanche vulnerable areas can be seen in summer. According to the compass, the slope is about 55°~60°, both to the south and to the north, to the northeast and to the southwest, and is perpendicular to the Palong Zangpo River. There is basically no vegetation cover on the surface. Close observation of the underlying soil is fine sand, easy to flow, there are more moraines on the surface, which can be judged as the remains of the vulnerable area of avalanche. After measuring at three points, each point five times, a total of 15 times, the soil temperature is about 23.5 degrees Celsius, but considering that the measurement time is in the afternoon, direct sunlight, high temperature, the result may not be accurate, the soil is relatively dry.

The effect of a sudden increase in surface temperature on avalanche excitation is quite significant. Changes in surface temperature can change the physical properties of the snow layer and cause avalanches to occur.

A sudden increase in precipitation is one of the most important factors that cause avalanches to occur. This can cause an increase in moisture in the snow layer, making it looser and wetter, and thus destabilizing. At the same time, snowfall is also included in precipitation, and increased precipitation can cause the gravity of the snow layer to rapidly increase beyond the strength of the snow layer, leading to avalanches.

According to the statistical data of the frequency and spatial distribution of avalanche disasters in the two study areas from 1980 to 2020, the slope, slope direction statistics, data fitting and scatter regression analysis of all the places where avalanche disasters occurred can be concluded as follows: Avalanche disasters in the two study areas occur more on the negative slope, that is, the southward distribution is more in summer, and the northward distribution is more in winter; When the slope is about 20°, the frequency of avalanche disasters begins to increase, when the slope is from 30° to

50°, the frequency of avalanche disasters reaches a peak at about 40°, and when the slope is above 60°, the frequency of avalanche disasters decreases.

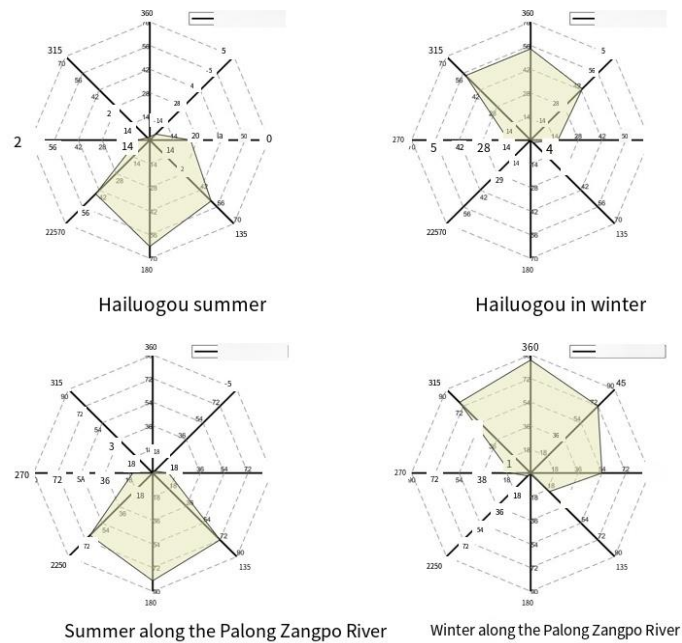

FIG. 4 Relationship between slope direction and avalanche disaster occurrence times in different seasons in the study area

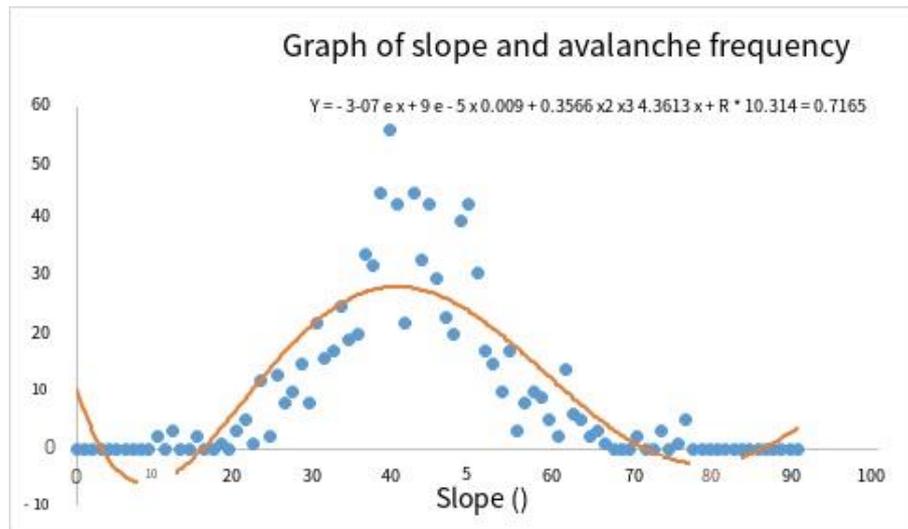

FIG. 5 Relationship between slope and cumulative avalanche disaster occurrence times in the study area

Based on the data set of monthly precipitation in China, the 8-day composite data set of global surface temperature/emissivity, and the statistical data of the number and spatial distribution of avalanche disasters in the two study areas from 1980 to 2020, all the days when the difference between precipitation and surface temperature and the average value of this month is within ±10mm/°C, and the cumulative number of avalanche disasters in all corresponding days are counted. The influence of precipitation anomaly and surface temperature anomaly on avalanche occurrence is obtained, and the following conclusions can be drawn: the sudden decrease of precipitation and surface temperature has little influence on avalanche occurrence, while the sudden increase of precipitation and surface temperature has great influence on avalanche occurrence. When the precipitation is 3mm higher than the average monthly precipitation, the number of avalanche disasters increases rapidly. When the surface temperature is 5°C higher than the average monthly precipitation, the number of avalanche disasters increases rapidly.

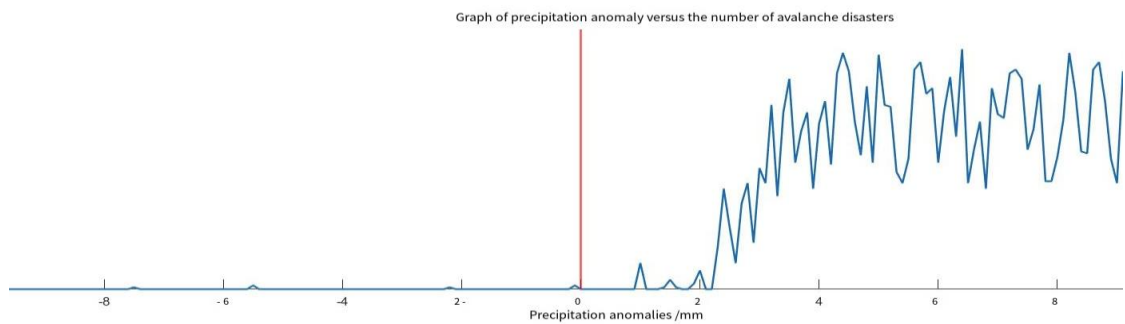

FIG. 6 Relationship between precipitation anomaly and avalanche disaster frequency

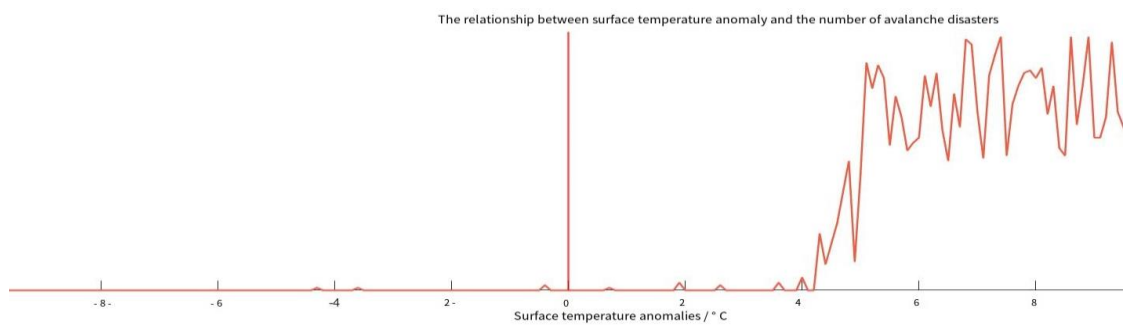

Figure 7. Relationship between surface temperature anomaly and the number of avalanche disasters

Therefore, based on the above avalanche model, field investigation results and statistics of avalanche disaster occurrence times, the threshold conditions for avalanch-prone areas can be obtained: the maximum monthly precipitation in summer is higher than 5mm, and the maximum surface temperature is greater than 0 degrees. In winter, the daily precipitation is higher than 5mm, while the surface temperature is greater than -10 degrees; The surface type is glacier, the slope is greater than 30° and less than 60°, and it is located on the negative slope, and the slope direction is 110°~250° in summer, 290°~360° and 0°~70° in winter. All pixels that meet the above conditions can be regarded as avalanch-prone areas.

**2.3.3 Extracting avalanch-prone areas based on threshold analysis**

According to the avalanche occurrence conditions in 3.2.2, the avalanche prone regions are extracted in PIE-Enginee, and the avalanche disaster prone regions of the two research areas during the 10 years from 2010 to 2019 are extracted. Since the threshold setting refers to the comprehensive results of avalanche occurrence model, field investigation and a large number of statistical data, the prediction accuracy is high, so this method can be used to produce products in a long series and a large range of avalanche prone areas.

**2.3.4 Visual consistency accuracy detection by the two methods**

The two methods were used to extract the avalanch-prone area of the Palong Zangbo River research area, and the consistency of the two methods in terms of shape, size, texture, style and spatial position was judged by visual interpretation, and the extraction accuracy of the two methods was evaluated.

**2.3.5 Spatial analysis of earthquake for avalanche disaster occurrence**

According to the statistical data of earthquake magnitude and spatial distribution in Hailuogou research area from 2000 to 2020, ArcMap was used to mark the location magnitude of earthquake disaster and the occurrence location of avalanche disaster at corresponding locations in Hailuogou research area, and the spatial correlation between earthquake disaster and avalanche disaster was analyzed and evaluated according to the spatial distribution of the two.

## 3. Research achievements

### 3.1 U-net Convolutional neural network extraction results

This study has generated a mature U-net convolutional neural network that can accurately identify avalanch-prone areas. Due to the limited amount of Gaofen-3 SAR data obtained, the trained convolutional neural network is used to identify the summer and winter images of 2020 along the Palong Zangpo River.

### 3.2 Avalanch-prone regions are extracted based on threshold analysis

In this study, the threshold analysis method was used to extract the interannual avalanch-prone regions in summer and autumn of the two study areas from 2010 to 2020. The blue region in the figure is the avalanch-prone region.

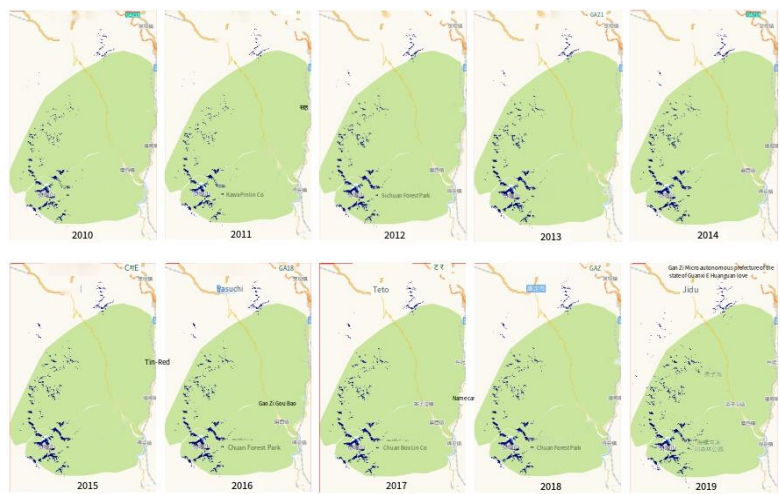

FIG. 8 Interannual time series of winter avalanche disaster prone areas in Hailuogou research area

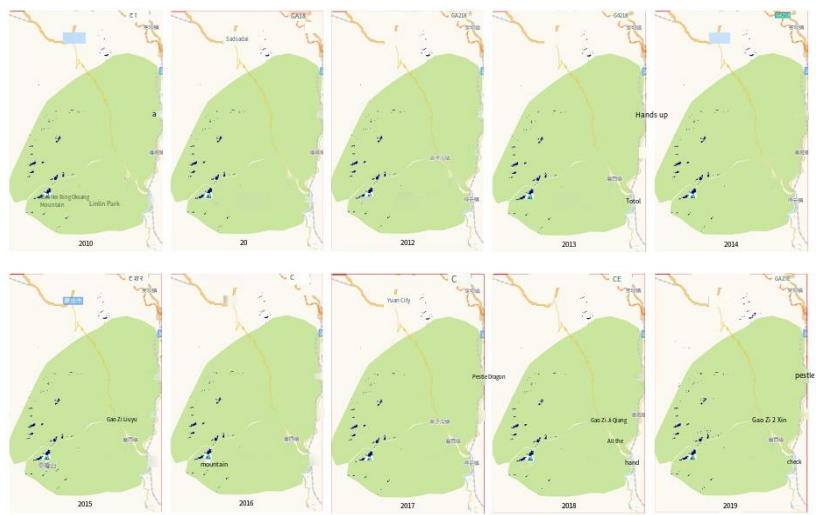

FIG. 9 Interannual time series of summer avalanche disaster prone area in Hailuogou Research area

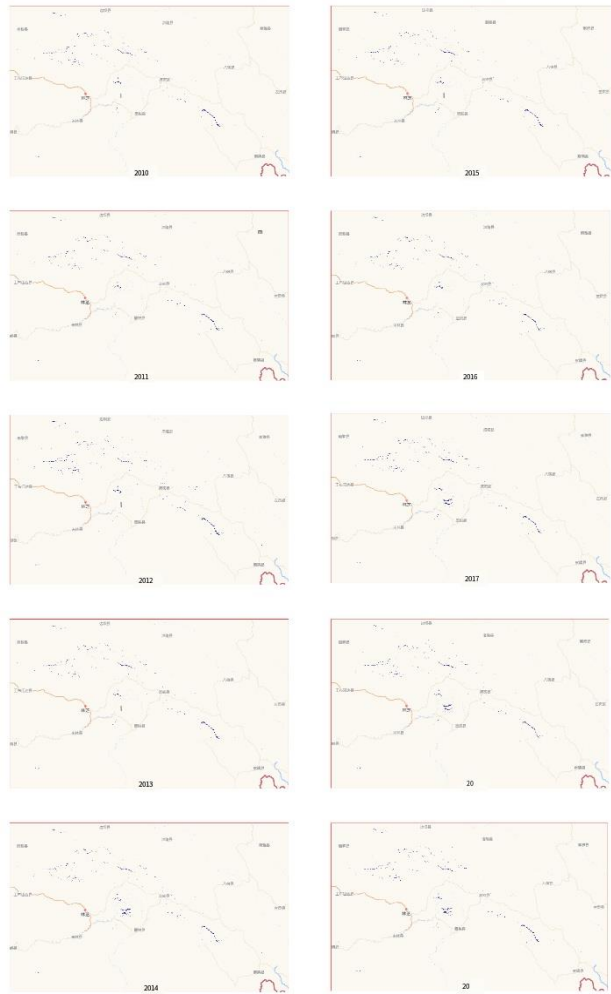

FIG. 10 Interannual time series of winter avalanche disaster prone area of Palong Zangpo River research area

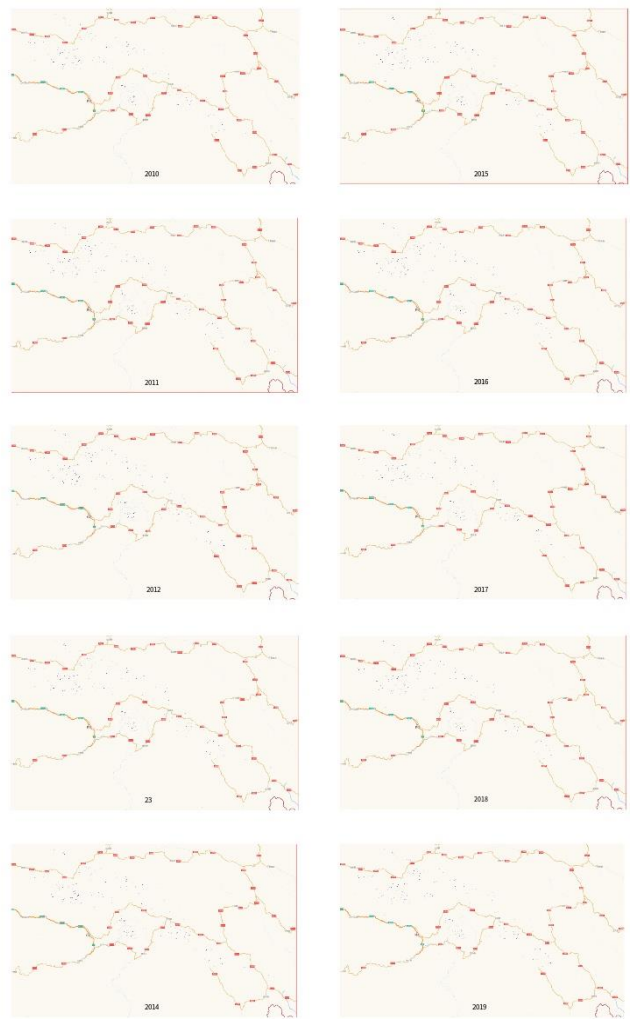

Figure 11 Interannual time series of summer avalanche disaster prone area of Palong Zangpo River coastal research area

From the figure, it can be found that the area prone to avalanche disaster in summer is smaller than that in winter in both study areas, and the area prone to avalanche gradually increases with the increase of years, which may be related to the melting of glaciers caused by global warming. The avalanche prone area of Hailuogou research area is mainly concentrated in the glaciers around Gongga Mountain, while the avalanche prone area of Palong Zangbu River research area is mainly dotted, which is strongly spatially correlated with the glacier distribution in the study area.

**3.3 Visual consistency accuracy test results of the two methods**

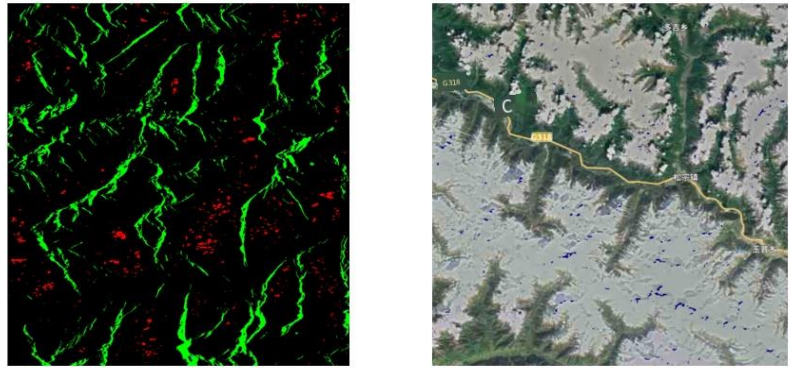

FIG. 12 Comparison of avalanch-prone areas in the Palong Zangbo River research area extracted by the two methods in summer

It can be found that in the study area along the Palong Zangpo River, the shape, size, texture and style of the avalanch-prone areas identified by the two methods have strong consistency, and they are scattered and scattered in the gullies on both sides. The area of each avalanch-prone area is small, and the avalanch-prone areas are distributed in clusters. However, the spatial consistency of the two methods is not very strong. The avalanche prone area extracted by U-net convolutional neural network basically covers the area extracted by threshold analysis. The reason may be that U-net convolutional neural network will misjudge some other disaster areas with similar texture. Therefore, it can be preliminarily determined that both of them have the ability to identify avalanch-prone areas, and can carry out basic extraction of avalanch-prone areas and production of long time series products.

**3.4 Results of spatial analysis of earthquake for avalanche disaster occurrence**

According to the statistical data of the number and spatial distribution of avalanche disasters in Hailuogou Research area from 1980 to 2020 and the statistical data of earthquake magnitude and spatial distribution from 2000 to 2020, the avalanche locations in Hailuogou Research area during the 20 years from 2000 to 2020 and the earthquake locations and magnitudes in the 16 days before each avalanche are calculated. The spatial distribution map is as follows:

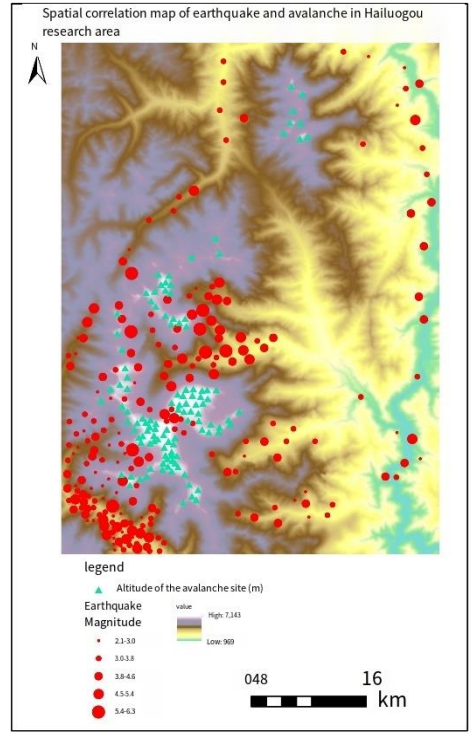

FIG. 13 Spatial correlation map of earthquake and avalanche in Hailuogou research area

There is a strong spatial correlation between earthquakes and avalanches in Hailuogou research area. Earthquakes are generally distributed near avalanch-prone areas, and the frequency of avalanches increases significantly with the increase of earthquake magnitude. This point can also be explained by the first law of geography, that is, "everything is related, but similar things are more closely related". The earthquake that occurred before the avalanche and the avalanche are close in space distance, so the two are more closely related.

Therefore, according to the above conclusions, when carrying out engineering construction and mountaineering activities, it is also necessary to consider the recent occurrence and magnitude of earthquakes near the relevant areas, and synthesize the distribution of avalanch-prone areas to assess the possible avalanche disaster risk.

## 4. Discussion and outlook

In this study, two methods, U-net convolutional neural network and threshold analysis, were used to extract avalanch-prone areas in the study area, and the accuracy of the two methods was verified, which proved that the two methods could be used to produce products in a long series and a large range of avalanch-prone areas, and further studied the spatial correlation between earthquake and avalanche disasters.

First of all, for U-net convolutional neural network, its training is based on the Gaofen-3 SAR L1A data product. Therefore, it is only suitable for the recognition of high-resolution SAR images, and pre-processing operations such as filtering should be performed before recognition. Therefore, if this method is used to produce avalanche vulnerable area products in a long time series and a large range, all high-resolution SAR data products in the whole year, winter and summer in a large spatial range are required, which has a large demand for data quality and quantity. In addition, some similar textures may be misjudged, and some landslide, debris flow and other areas may be wrongly identified as avalanch-prone areas. Therefore, further analysis can be made on the texture, shape, size and other characteristics of avalanch-prone areas and other disaster-prone areas to distinguish the detailed differences between the two, and further analysis can be carried out in combination with other types of remote sensing image differences. Continue to use more sources of remote sensing images to further train the network and improve the recognition accuracy.

Secondly, for the threshold analysis method, according to the conditions of avalanche occurrence model, field investigation and years of statistical data, it can be judged that the determination of threshold conditions is basically accurate, and it can initially reflect the excitation effect of slope, slope direction, precipitation and surface temperature anomalies on avalanche. However, this method still has some shortcomings, mainly reflected in the lack of prior knowledge. For the occurrence mechanism and model of avalanches, the current understanding is still insufficient. It is necessary to combine the physical structure inside the snow layer and the adhesion relationship between snow particles, and add more abundant climatic conditions, such as anomalies of surface humidity, air temperature and snow depth. Simulation experiments under different conditions are carried out to further accumulate prior knowledge of avalanche occurrence conditions and build prior knowledge base of avalanche occurrence, so as to further identify avalanch-prone areas more accurately according to various restrictive conditions.

Then there is the impact of earthquakes on avalanches. Earthquakes can trigger, aggravate and even cause huge avalanches under special conditions; It can also change the stability of the avalanche material, so that it is out of balance; Earthquake can accelerate the avalanche material, increase the intensity of the avalanche damage; And under some special conditions, earthquake can cause avalanche group, that is, multiple avalanches occur at the same time; Earthquakes can also trigger ice collapses. Aiming at the spatial distribution relationship between earthquakes and avalanches, the relationship model between earthquakes of different distance and magnitude and the number and degree of avalanches nearby can be further established, and combined with the method of threshold analysis and extraction can further increase the prior knowledge and improve the accuracy of prediction.

Finally, after improving the accuracy of the two methods, the accuracy of the two methods is first determined in the study area, and the accuracy of the judgment and recognition results is further combined with field investigation. After the higher precision method was determined, the two methods were further used to extract avalanch-prone areas in southwest China (including Sichuan Province, Yunnan Province and Tibet Autonomous Region), and produce long time series of interannual products.

These products can provide early warning of avalanche disasters in southwest China, and guarantee site selection, construction safety and later maintenance for the "14th Five-Year Plan" national large-scale key projects such as the Sichuan-Tibet Railway and Sichuan-Tibet High-speed crossing the Hengduan Mountains and the Himalayas. At the same time, there are many alpine expeditions carried out in Sichuan-Tibet areas every year, and the products can help these activities predict avalanche disasters in advance and help them make preparations to reduce casualties in alpine expeditions.

## Acknowledgements

Thanks to Teacher Kang Zhizhong for providing the overall thought guidance and theory teaching of each step for this research. Meanwhile, Teacher Kang Zhizhong also provided the

Gaofen-3 SAR L1A data products and SRTM30mDEM data products for the Palong Zangbo River Coast Research Area for this research, and provided the server for U-net convolutional neural network training; In addition, I would like to thank Mr. Geng Yandong for his practical guidance in U-net convolutional neural network training, and Mr. Xu Xiaojian for his guidance in later data processing. Meanwhile, I would like to thank Mr. Yan Kai and Mr. Zhong Run for their theoretical and practical guidance on Pie-Engine in remote sensing principle and application courses, which helps this research find ideas and breakthroughs in threshold analysis. We also want to thank all the members of this group and the other members who have contributed to this research; The National Tibetan Plateau Scientific Data Center and the National Seismic Science Data Center for their data support; Finally, I would like to pay tribute to all the front-line personnel who struggle in Sichuan-Tibet Railway and high-speed projects, and the climbers who bravely climb the peak and constantly exceed the limit. I hope that the products of this research can add a guarantee for your safety.

## 【 Reference 】